\documentclass[conference]{IEEEtran}
\IEEEoverridecommandlockouts

\usepackage{cite}
\usepackage{amsmath,amssymb,amsfonts}
\usepackage{subcaption}
\usepackage{algorithmic}
\usepackage{graphicx}
\usepackage{textcomp}
\usepackage{xcolor}
\usepackage{braket}

\usepackage{amsfonts}
\usepackage{times}
\usepackage{latexsym}
\usepackage{bm}
\usepackage{dsfont}
\usepackage{amssymb}
\usepackage{amsmath}
\usepackage{cite}
\usepackage{verbatim}
\usepackage{algorithm}
\usepackage{algorithmic}
\usepackage{svg}

\newcommand{\EM}{EM}

\def\ml{ML}
\def\zne{ZNE}
\def\pec{PEC}
\def\qem{QEM}

\def\ber{BER}

\def\bb0{{\mathbb{0}}}

\def\bb{{\mathbf{b}}}

\def\bx{{\mathbf{x}}}
\def\by{{\mathbf{y}}}

\def\b0{{\mathbf{0}}}

\def\bX{{\mathbf{X}}}
\def\bY{{\mathbf{Y}}}
\def\bZ{{\mathbf{Z}}}

\def\bbE{{\mathbb{E}}}

\def\bbP{{\mathbb{P}}}

\def\cG{\mathcal{G}}

\def\cL{\mathcal{L}}

\def\cW{\mathcal{W}}

\def\cY{\mathcal{Y}}
\def\cZ{\mathcal{Z}}

\def\sf0{{\mathsf{0}}}

\def\balpha{\boldsymbol{\alpha}}

\def\bvarepsilon{\boldsymbol{\varepsilon}}

\def\beta{\boldsymbol{\eta}}
\def\btheta{\boldsymbol{\theta}}

\def\BibTeX{{\rm B\kern-.05em{\sc i\kern-.025em b}\kern-.08em
    T\kern-.1667em\lower.7ex\hbox{E}\kern-.125emX}}

\DeclareMathOperator*{\argmax}{arg\,max}

\begin{document}

\title{Statistical Signal Processing \\for Quantum Error Mitigation\\
\thanks{The project is funded by the U.S. Department of Energy, Advanced Scientific Computing Research, under contract number DE-SC0025384, and the Portuguese Foundation for Science and Technology (FCT), under grant UID/50008}}

\author{
\IEEEauthorblockN{Kausthubh Chandramouli}
\IEEEauthorblockA{\textit{North Carolina State University} \\
Raleigh, NC, USA \\
kprabha@ncsu.edu}
\and
\IEEEauthorblockN{Kelly Mae Allen}
\IEEEauthorblockA{\textit{North Carolina State University} \\
Raleigh, NC, USA \\
kmallen5@ncsu.edu}
\and
\IEEEauthorblockN{Christopher Mori}
\IEEEauthorblockA{\textit{North Carolina State University} \\
Raleigh, NC, USA \\
cpmori@ncsu.edu}
\and[\hfill\mbox{}\par\mbox{}\hfill]
\IEEEauthorblockN{Dror Baron}
\IEEEauthorblockA{\textit{North Carolina State University } \\
Raleigh, NC, USA \\
dzbaron@ncsu.edu}
\and
\IEEEauthorblockN{M\'ario A. T. Figueiredo}
\IEEEauthorblockA{\textit{Instituto Superior T\'ecnico} \\
Lisbon, Portugal \\
mario.figueiredo@tecnico.ulisboa.pt}
}

\maketitle

\begin{abstract}
In the noisy intermediate-scale quantum (NISQ) era, quantum error mitigation (QEM) is essential for producing reliable outputs from quantum circuits. We present a statistical signal processing approach to QEM that estimates the most likely noiseless outputs from noisy quantum measurements. Our model assumes that circuit depth is sufficient for depolarizing noise, producing corrupted observations that resemble a uniform distribution alongside classical bit-flip errors from readout. Our method consists of two steps: a filtering stage that discards uninformative depolarizing noise and an expectation-maximization (EM) algorithm that computes a maximum likelihood (ML) estimate over the remaining data. We demonstrate the effectiveness of this approach on small-qubit systems using circuit simulations in Qiskit and IBM quantum processing unit (QPU) data, and compare its performance to contemporary statistical QEM techniques. We also show that our method scales to larger qubit counts using synthetically generated data consistent with our noise model. These results suggest that principled statistical methods can offer scalable and interpretable solutions for quantum error mitigation in realistic NISQ settings.
\end{abstract}


\section{Introduction}
The noisy intermediate-scale quantum (NISQ) era of quantum computing requires strategies beyond the quantum circuit to ensure accurate results~\cite{NISQ}. Quantum error mitigation ({\qem}) works by mitigating or reducing noise by post-processing circuit outputs using classical computation to extrapolate or estimate ideal results~\cite{QEM}. This work adopts a post-processing {\qem} approach, leveraging assumed characteristics of arbitrary quantum circuits to infer noiseless outputs. Our work is agnostic to the specific details of a quantum circuit and implicitly estimates the density function of an output quantum state in the computational basis. 

Some state-of-the-art methods for error mitigation have demonstrated favorable results in reducing the effect of circuit noise already. Zero-noise extrapolation ({\zne}) and probabilistic error-cancellation ({\pec}) have been extensively studied in the context of short-depth quantum circuits and have proven effective in both theoretical and experimental settings by extrapolating a noiseless quantum state~\cite{Temme_EM,Endo_EM,Song_2019}. Recent complexity analyses suggest that the number of circuit repetitions, shots, for accurate mitigation grows with circuit depth~\cite{Quek_Bounds}. 

We group methods that apply inference and learning-based strategies to mitigate errors as statistical {\qem} techniques~\cite{Clifford_EM, bayesian_unfolding, QMV, hammer, qbeep, M3}. Clifford data regression builds predictive models by training on classically simulated circuits and fitting noisy results to the learned regression function~\cite{Clifford_EM}. However, because this method depends on classical simulation, it becomes infeasible as circuit complexity and qubit count increase. Majority-vote decoding optimally finds the maximum likelihood (\ml) solution for circuits with a single valid result~\cite{QMV}. Under the simplified assumption that the quantum algorithm has one correct solution, it turns out that majority-vote is optimal. The expectation-maximization algorithm refines output estimates in systems affected primarily by bit-flip readout errors~\cite{bayesian_unfolding}. This approach is effective but only handles readout errors instead of quantum gate errors. Our approach builds on these statistical ideas and extends them to arbitrary quantum circuits with gate errors and multiple valid outputs. We require only a coarse understanding of the structure of erroneous results in order to identify the set of valid solutions for a given quantum circuit. This strategy draws on principles from classical signal processing, where similar statistical estimation problems have been studied extensively.

Conventional signal processing often uses statistical techniques to estimate a signal among noise. Maximizing a likelihood function is a common strategy for estimation~\cite{VanTrees,prorakis}. When a likelihood function has multiple local maxima, schemes such as the expectation-maximization ({\EM}) algorithm iteratively search for one solution~\cite{EM_original}. The {\EM} approach extends to mixture models where multiple unknown components contribute to the observed data, and the number of components may also be unspecified~\cite{Mario_K}. Quantum problems, however, often have additional complexity that complicates direct application of these methods.

Signal recovery criteria depend on the underlying noise statistics. For quantum circuits, depolarizing noise that nearly uniformly distributes measurements across all possible observable bit strings is the worst-case scenario, but it is not necessarily uncommon with deep circuits~\cite{Wilde}. The depolarizing noise makes estimation difficult as it has little exploitable structure. Classical filtering techniques tackle this type of noise by removing outliers and suppressing uniformly distributed noise that carries little useful information~\cite{Yang,Ramaswamy,DBSCAN}.

Building on estimation and filtering, we develop a method for estimating quantum circuit outputs from noisy measurements. Our approach combines {\qem}, statistical signal processing, and classical filtering to identify the most likely states underlying a given set of observations. We assume that ({i}) the relevant data correspond to binary vectors; ({ii}) quantum circuits may yield multiple valid outputs; and ({iii}) the underlying errors include both depolarizing behavior and bit-flips. We introduce a simple filtering strategy to discard noisy observations consistent with depolarizing noise. We then adapt the {\EM} algorithm to binary output distributions. Finally, we demonstrate this pipeline on simulated quantum circuits to evaluate its performance.

We assert that an ideal {\qem} system should scale to quantum circuits with large qubit counts, which implies that its computational complexity must be polynomial in the number of qubits, $n$. A {\qem} system must also recover valid outputs that are never directly observed due to extreme noise, as in algorithms like Bernstein-Vazirani~\cite{Bernstein}. Unlike such single-solution cases, many quantum algorithms yield multiple valid outputs even in the absence of noise. A robust {\qem} system should accommodate a growing number of correct outputs, $K$. Furthermore, quantum gate errors, especially in mid-circuit, can propagate and corrupt many qubits, often resulting in output distributions consistent with heavy depolarizing noise. This behavior, observed in practice where 80--90\% of shots appear random, must be handled effectively. Measurement crosstalk and readout errors, which can reach probabilities up to 0.2 under non-ideal conditions, must also be addressed. Our presented scheme addresses the challenges natively through our noise model (Sec.~\ref{sec:noise_model}), likelihood function (Sec.~\ref{sec:likelihood_function}), and algorithmic procedure (Sec.~\ref{sec:procedure}). Our experimental results demonstrate our approach's  effectiveness (Sec.~\ref{sec:experimental}).

\section{Problem formulation}
\subsection{Noise model}\label{sec:noise_model}
We model a noise process acting on a qubit in state $\rho$. The process naturally extends to an ensemble of $n$ qubits, with the joint state represented by $\rho^{\otimes n}$. Our model acts independently on each qubit. The process transforms a state $\rho$ to the maximally mixed state $\pi$ with probability $p$ and leaves the state unchanged with probability $(1-p)$. The maximally mixed state for a single qubit is the equal superposition of $\ket{0}$ and $\ket{1}$. The state after the noise process is
\begin{equation}\label{noisy_state}
        \rho \rightarrow(1-p)\rho + p\pi \,,
    \end{equation}
which corresponds to standard depolarizing noise. Depolarizing noise models average noise effects in deep circuits \cite{urbanek2021mitig}, and represents a worst-case scenario, as it removes all correlations with the input state, providing no information for recovery~\cite{Wilde}. Further, employing twirling techniques such as Pauli twirling \cite{bennett1996mixed, cai2019constructing} converts arbitrary noise channels into Pauli channels that align with our depolarizing noise assumption.

We further assume that readout after measurement is subject to a bit-flip process. This process is strictly classical. With some probability, $\varepsilon$, a classical bit is flipped from $0$ to $1$ or $1$ to $0$. With probability $(1-\varepsilon)$ the bit is not affected by readout error. The overall process describing the evolution of the classical bit $b$ to $b'$ after readout,
\begin{equation}\label{eq:bit_flip}
    b' = 
    \begin{cases}
    \overline{b}, & \text{with probability } \varepsilon \\
    b, & \text{with probability } 1 - \varepsilon
    \end{cases}
\end{equation}
captures arbitrary readout errors and, importantly, assumes the crossover probabilities of $0$ to $1$ and $1$ to $0$ are the same.

\subsection{Likelihood function} \label{sec:likelihood_function}
If depolarizing noise is removed by some pre-processing step, we only model remaining bit-flip errors. Sec.~\ref{filtering} outlines a process to remove depolarizing noise. The noise model (Sec.~\ref{sec:noise_model}) extends to each qubit. Each is affected independently and the bit-flip probabilities $\varepsilon_i$ may vary across qubits. That is, $\boldsymbol{\varepsilon} = \{\varepsilon_1, \varepsilon_2, \ldots, \varepsilon_n\}$, where each $\varepsilon_i$ may differ from the others. 

The quantum circuit is executed $S$ times (shots), producing a dataset $\mathcal{Y} = \{ \by^{(1)}, \ldots, \by^{(S)} \}$ of independent noisy outputs, where $\by^{(i)}\in\{0,1\}^n$. Our goal is to recover the noiseless results of the quantum circuit, assumed to lie in a set of $K$ binary vectors $\bX = \{\bx_1, \bx_2, \ldots, \bx_K\}$, where $K$ is unknown. Each vector $\bx_k , \in\{0,1\}^n$ is associated with a mixing weight $\alpha_k$, with the full set of weights denoted by $\balpha = (\alpha_1, \ldots, \alpha_K)$. A complete parameter set is $\btheta = \{\bX,\balpha,\boldsymbol{\varepsilon}\}$.

To express the likelihood, we introduce a latent variable $\bZ \in \{0, 1\}^K$, a one-hot vector indicating which $\bx_k$ generated a given noisy output $\bY$. The likelihood that $\bx_k$ generates a particular observation $\by$ is 
\begin{equation}\label{PYZ_IDD}
\bbP[\bY \! =\! \by \mid Z_k \! =\! 1] 
=  \prod\limits_{j=1}^{n} \varepsilon_j^{y_{j} \oplus x_{kj}} (1 - \varepsilon_j)^{1 - (y_{j} \oplus x_{kj})},
\end{equation}
where $\oplus$ denotes modulo-2 sum (\textit{i.e.}, the logical XOR operation), $y_j$ is the $j$-th component of $\by$, and $x_{kj}$ is the $j$-th component of $\bx_k$. Since $\mathbb{P}[Z_k=1] = \alpha_k$, marginalizing over $\bZ$ yields the finite mixture likelihood
\begin{align}\label{prob_Y_IDD}
    \bbP[\bY = \by \mid & \bX, \balpha, \varepsilon]  \\ \notag
    & = \sum_{k=1}^K \bbP[\bY = \by \mid Z_k = 1, \bX, \bvarepsilon] \; \bbP(Z_k = 1 | \balpha)  \\
    & = \sum_{k=1}^K \alpha_k \prod\limits_{j=1}^{n} \varepsilon_j^{ y_{j} \oplus x_{kj}} (1 - \varepsilon_j)^{1 - (y_{j} \oplus x_{kj})} \notag \,,
\end{align}
which represents the probability of observing $\by$ given all parameters in $\boldsymbol{\theta}$. We then have the log-likelihood over the $S$ observed samples,
\begin{equation}\label{log_likelihood_IDD}
    \cL(\boldsymbol{\theta}; \mathcal{Y}) = \sum_{i=1}^S \log \sum_{k=1}^K \alpha_k \prod\limits_{j=1}^{n} \varepsilon_j^{y_{ij} \oplus x_{kj}} (1 - \varepsilon_j)^{1 - (y_{ij} \oplus x_{kj})},
\end{equation}
where $y_{ij}$ denotes the $j$-th component of $\by^{(i)}$. $\cL(\boldsymbol{\theta}; \mathcal{Y})$ expresses how likely a particular parameter setting is. Finally, the maximum likelihood {\ml} estimate  is given by
\begin{equation}\label{ML_estimator}
        \widehat{\btheta} = \argmax_{\btheta} \mathcal{L}(\btheta;\mathcal{Y}).
\end{equation}
As in other mixture models, this optimization cannot be solved in closed form. Instead, we apply the expectation-maximization (EM) algorithm~\cite{EM_original}, which converges to a local maximum of $\mathcal{L}(\btheta;\mathcal{Y})$. The derivation of the EM algorithm for our problem is presented in Sec.~\ref{sec_EM}.

\subsection{Estimating $K$: Modified Likelihood Function}
As is well known, if $K$ is unknown, the likelihood function alone cannot be used as a selection criterion, as it is an increasing function of $K$~\cite{Mario_K}. We thus adopt a modified version of the log-likelihood function based on the minimum message length (MML) model selection approach~\cite{Mario_K}, which is given by 
\begin{align}\label{eq:likelihoodMML}
    \cL_{\text{MML}}(\btheta; \cY) & = -\frac{K_{nz}}{2}\log\frac{S}{12} 
    - \frac{K_{nz}n + K_{nz}}{2}  \\ \notag
    & \;\;\; \;\;  + \cL(\btheta; \cY) - \frac{d}{2} \sum_{k:\,\alpha_k>0} \log \frac{S\alpha_k}{12}
\end{align}
where $\cL(\btheta; \cY)$ is the original log-likelihood function in \eqref{log_likelihood_IDD}, $K_{nz}$ is the number of components with non-zero mixing probabilities ($\alpha_k > 0$), and $d$ is the number of parameters specifying each component; in our model, $d=n$, the dimensionality of each $\bx_k$.

The MML model we use originated from studies of Gaussian mixtures~\cite{Mario_K}. However, some aspects of our estimation procedure differ from the Gaussian approach. For example, the log-likelihood function~\eqref{eq:likelihoodMML} should account for the coding length required to describe $K$ solutions of length $n$ bits, and we should replace the term $\frac{N_{nz}n}{2}$ with $K_{nz}n$. MML characterization is ongoing work. The current formulation~\eqref{eq:likelihoodMML} works sufficiently well (Sec.~\ref{sec:experimental}) so we focused on the core {\EM} algorithm without significant modification.

\section{Algorithmic procedure}\label{sec:procedure}

Estimating the noiseless outputs of some quantum circuit has two parts. The first part, described by Algorithm~\ref{alg:junk_shots_filter}, removes depolarizing noise from the observed data set. The second part, described by Algorithm~\ref{alg:EM}, performs the {\EM} algorithm on the filtered data set to extract the {\ml} solution to the remaining data.

\subsection{Filtering}\label{filtering}
Removing the shots that resemble depolarizing noise from the data set is the first step to estimating the noiseless quantum result. Removal of the uniformly spread depolarizing noise yields data more consistent with the bit-flip-only model~\eqref{eq:bit_flip}. Algorithm~\ref{alg:junk_shots_filter} outlines our process to remove depolarizing noise. We use a density-based idea by constructing a threshold over local Hamming neighborhoods. Given the set of shots of length $S$, the number of bits $n$, and a tuneable threshold parameter $\eta$, we filter the data. We first set some basic threshold that approximates the result of total uniform noise.

\begin{algorithm}[h]
    \caption{Depolarization filter}
    \label{alg:junk_shots_filter}
    \begin{algorithmic}[1]
        \REQUIRE Binary shot dataset of size $S$, threshold $T$
%        \STATE \st{Estimate expected frequency under uniform noise \\$\lambda = S/2^n$}
        \STATE Count bit-string frequencies and 1-Hamming bit-string neighbors
        \STATE Suppress low-support entries below threshold $T$
        \STATE Reconstruct dataset from remaining high-support entries
        \RETURN Filtered shot dataset
    \end{algorithmic}
\end{algorithm}

We then count the frequency of every observed bit-string $x$ and the number of 1-Hamming distance bit-strings from $x$. The frequency and 1-Hamming neighbors form a count, $f(x)$, for every bit-string, $x$, in our shots $\cY$. With $S$ shots and $2^n$ possible bit vectors, the expected uniform frequency is $\lambda = S/2^n$.  
Because $f(x)$ sums the counts of $n+1$ possible bit strings, its expected value obeys, $E[f(X)]=\lambda(n+1)$.
We calculate a threshold, $T$, a function of $\lambda$ and $n$; it should be somewhat larger than $\lambda(n+1)$. Finally, we process the counts, $(f(x))_{x\in\cY}$, and remove shots, $x \in \cY$, whose corresponding
counts, $f(x)$, are smaller than $T$.
Details of our filtering approach appear in  Algorithm~\ref{alg:junk_shots_filter}.

\subsection{Expectation maximization (\EM) algorithm} \label{sec_EM}
The {\EM} algorithm is an iterative method for finding {\ml} estimates of parameters in models with latent (unobserved) variables. In the model described in Sec.~\ref{sec:likelihood_function}, the latent variables are $\mathcal{Z} = \{\bZ^{(1)},\ldots,\bZ^{(S)}\}$, where each $\bZ^{(i)}$ indicates the mixture component that produced the corresponding observation $\by^{(i)}$. The complete log-likelihood (the one that is maximized if $K$ and $\mathcal{Z}$ are known) is
  \begin{align}\label{complete_log_likelihood}
        \cL_c(\btheta; \cY, \cZ) 
        &= \sum_{i=1}^S \sum_{k=1}^K Z_{ik} \log \alpha_k \notag \\
        \quad+ \sum_{i=1}^S \sum_{k=1}^K Z_{ik} \log &\prod\limits_{j=1}^{n} \varepsilon_j^{y_{ij} \oplus x_{kj} } (1 - \varepsilon_j)^{1 - (y_{ij} \oplus x_{kj})}.
\end{align}

The classic {\EM} algorithm consists of two steps: the expectation step (E-step), in which the expectation with respect to $\mathcal{Z}$ of the complete log-likelihood is computed based on current parameter estimates, and the maximization step (M-step), which updates all parameter estimates by maximizing this expected complete log-likelihood~\cite{EM_original}. Our version maximizes $\cL_{\text{MML}}(\btheta; \cY)$ instead of $\cL(\btheta; \cY)$, but this only affects the M-step. Algorithm~\ref{alg:EM} describes the full EM algorithm.

\subsubsection{E-step}
The expectation of the complete log-likelihood function given the current parameter estimates at iteration $t$ and the observations $\mathcal{Y}$ leads to the so-called $Q$-function
\begin{equation}\label{Estep}
        Q(\btheta, \widehat{\btheta}^{(t)}) = \bbE_{\cZ}[\cL_c(\btheta; \cY, \cZ) | \cY, \widehat{\btheta}^{(t)}] ,
    \end{equation}
where $\mathbb{E}_{\mathcal{Z}}$ denotes expectation with respect to $\mathcal{Z}$. Because the complete log-likelihood function is linear with respect to the latent variables, 
\begin{equation}
    Q(\btheta, \widehat{\btheta}^{(t)}) = \bbE_{\cZ}[\cL_c(\btheta; \cY, \cZ) | \cY, \widehat{\btheta}^{(t)}] = \cL_c(\btheta; \cY, \cW) \,,
\end{equation}
where $\cW = \bbE[\cZ | \cY, \widehat{\btheta}^{(t)}]$ is the conditional expectation of the latent variables given the observations and the current parameter estimates. Since each $\bZ^{(i)}$ depends only on the corresponding $\by_i$ and each $Z_{ik}$ is a binary variable, a simple application of Bayes rule yields
\begin{align}\label{EStepIDD}
    W_{ik} &= \bbE[Z_{ik} | \by_i, \widehat{\btheta}] = \bbP[Z_{ik} = 1 | \by_i, \widehat{\btheta}] \nonumber\\
    &= \frac{\widehat{\alpha}_k \prod\limits_{j=1}^{n} \hat\varepsilon_j^{ y_{ij} \oplus \widehat{x}_{kj}} (1 - \hat{\varepsilon}_j)^{1 - (y_{ij} \oplus \widehat{x}_{kj})}}{\sum\limits_{l=1}^{K} \widehat{\alpha}_l \prod\limits_{j=1}^{n} \hat\varepsilon_j^{y_{ij} \oplus \widehat{x}_{lj}} (1 - \hat{\varepsilon}_j)^{1 - (y_{ij} \oplus \widehat{x}_{lj})}}.
\end{align}

\subsubsection{M-step}
In this step, we update the parameter estimates by maximizing $\cL_{\text{MML}}(\btheta; \cY)$ with $Q(\btheta, \widehat{\btheta}^{(t)})$ replacing $\cL(\btheta; \cY)$~\eqref{eq:likelihoodMML}, which we denote by $\cL_{\text{MML}}(\btheta; \cY, \cW)$. 

The maximization of $\cL_{\text{MML}}(\btheta; \cY, \cW)$ with respect to $\balpha$ is independent of the other parameters and leads to 
\begin{align}\label{eq:alphaAutoK}
    \widehat{\alpha}_k^{(t+1)} = \frac{\max\biggl\{0,\biggl(\displaystyle \sum_{i=1}^{S}W_{ik}\biggr)-\frac{n}{2}\biggl\}}{\displaystyle \sum_{l=1}^{K}\max\biggl\{0,\biggl(\displaystyle\sum_{i=1}^{S}W_{il}\biggr)-\frac{n}{2}\biggl\}} .
\end{align}
This step effectively performs component annihilation (whenever some $\widehat{\alpha}_k^{(t+1)} = 0$), thus being an explicit rule for moving from the current number of non-zero probability components $K_{nz}$ to a smaller one.

To maximize with respect to $\bX$, consider $\cL_c(\btheta; \cY, \cW)$ after dropping the $\balpha$ term, denoting it by $\cG(\bX, \bvarepsilon)$:
\begin{align}\label{GXe}
    \cG(\bX, \bvarepsilon) = & \sum_{i=1}^S \sum_{k=1}^K W_{ik} \sum_{j=1}^{n} \biggl( (y_{ij}\oplus x_{kj}) \log \varepsilon_j \\ \notag
    &  + (1 - y_{ij} \oplus x_{kj}) \log (1 - \varepsilon_j)\biggr) \\ \notag
    & \hspace{-1.6cm} = \sum_{j=1}^n \biggl(S\log(1 - \varepsilon_j) + \log\frac{\varepsilon_j}{1-\varepsilon_j} \sum_{i=1}^{S} \sum_{k=1}^{K} W_{ik} ( y_{ij} \oplus x_{kj}) \biggr).
\end{align} 
The maximization with respect to $\bX$ is independent of $\bvarepsilon$; it is also decoupled across $\bx_1, \ldots, \bx_K$ and decoupled across the $n$ dimension. Assuming $\varepsilon_j < 0.5$ implies that $\log\frac{\varepsilon_j}{1-\varepsilon_j} < 0$, thus
\begin{align}\label{eq_updateX}
\widehat{x}_{kj}^{(t+1)} \! &= \arg\!\min_{x \in \{0,1\}} \sum_{i=1}^S W_{ik} (y_{ij} \oplus x) \\ \notag
&= H\Bigl( \sum_{i=1}^S W_{ik} (2 y_{ij}-1) \Bigr)\,,
\end{align}
where $H$ is the Heaviside function: $H(u)=1$, if $u\geq0$, and $H(u)=0,$ if $u<0$. 

Finally, the maximization with respect to $\bvarepsilon$ leads to 
\begin{equation}\label{eq:updateEpsIDD}
    \widehat{\varepsilon}_j^{(t+1)} = \frac{1}{S} \sum\limits_{i=1}^{S} \sum\limits_{k=1}^{K} W_{ik} ( y_{ij} \oplus \widehat{x}_{kj}^{(t+1)}).
\end{equation}

\begin{algorithm}[t]
    \caption{Expectation Maximization Algorithm}
    \label{alg:EM}
    \begin{algorithmic}[1]
        \REQUIRE Observations $\by^{(1)}, \dots, \by^{(S)}$
        \REQUIRE $K_{\text{min}}$\,, $K_{\text{max}}$ (upper and lower limits on $K$)
        \REQUIRE Initial estimates $\widehat{\bx}^{(0)}_1,..., \widehat{\bx}^{(0)}_{K_{\text{max}}}$, $\widehat{\alpha}^{(0)}_1, \dots, \widehat{\alpha}^{(0)}_{K_{max}}$, $\widehat{\bvarepsilon}^{(0)}$, which together form $\widehat{\btheta}^{(0)}$, and a stopping threshold $\delta$
        \ENSURE Optimal parameter estimates $\widehat{\btheta}_\text{{best}}$
        \STATE $t \gets 0$, $K_\text{{nz}} \gets K_\text{{max}}$, $\cL_\text{{min}} \gets + \infty$
        \WHILE{$K_\text{{nz}} \ge K_\text{{min}}$}
            \REPEAT
                \STATE $t \gets t + 1$
                \FOR{$k = 1$ \TO $K_{\text{max}}$} 
                    \STATE Perform E-step, obtaining $W_{ik}$,~\eqref{EStepIDD}
                    \STATE Compute $\widehat{\alpha}_k^{(t+1)}$,~\eqref{eq:alphaAutoK}
                    \IF{$\widehat{\alpha}_k > 0$}
                        \STATE Compute $\widehat{\bx}_{k}{}^{(t+1)}$,~\eqref{eq_updateX}
                        \vspace{.3ex}
                        \STATE Compute $\widehat{\bvarepsilon}^{(t+1)}$,~\eqref{eq:updateEpsIDD}
                    \ELSE
                        \STATE $K_{\text{nz}} \gets K_{\text{nz}} - 1$ 
                    \ENDIF
                \ENDFOR
            \STATE Compute $\cL_{\text{MML}}(\widehat{\btheta}^{(t)};\mathcal{Y})$,~\eqref{eq:likelihoodMML}
            \UNTIL{$\cL_{\text{MML}}^{(t-1)} - \cL_{\text{MML}}^{(t)} < \delta \left| \cL_{\text{MML}}^{(t-1)} \right| $}
            \IF{$\cL_{\text{MML}}(\widehat{\btheta}^{(t)};\mathcal{Y}) \le \cL_{min}$}
                \STATE $\cL_{\text{min}} \gets \cL_{\text{MML}}(\widehat{\btheta}^{(t)};\mathcal{Y})$
                \STATE $\widehat{\btheta}_{\text{best}} = \widehat{\btheta}^{(t)}$
            \ENDIF
            \STATE $k^{*} = \arg \min_{k} \{\widehat{\alpha}_k > 0\}$\,, $\widehat{\alpha}_{k^*} \gets 0$\,, $K_{nz} \gets K_{nz} - 1$
        \ENDWHILE
        \RETURN Final parameter estimates $\widehat{\btheta}_{\text{best}}$
    \end{algorithmic}
\end{algorithm}

\section{Experimental results}\label{sec:experimental}
Experimental results are derived using simulated IBM hardware in Qiskit~\cite{qiskit2024} and data from the IBM Brisbane QPU for small qubit counts. For systems where qubit counts are too large to simulate or run on hardware, we use synthetically generated data to demonstrate the accuracy of our algorithm.

We study quantum circuits that output measured binary bit-strings. We interpret errors as individual bits being incorrect and use the Hamming distance to quantify differences between bit-strings. Our experiments aim to identify a collection of $K$ bit-strings, each length $n$ bits. We use the bit error rate ({\ber}) to evaluate the total error across the set.

We compute {\ber} by comparing the estimated set, $\widehat{\bX}$, and true set, $\bX$, as follows: at each step, we pair two unmatched vectors, one from each set, that are closest in Hamming distance and then add their distance to a running total. Once paired, we exclude the vector from further matches. This process continues until either set runs out of vectors to match. In closed form, the bit error rate ({\ber}),
\begin{equation}\label{bit_error_rate}
\text{\ber} = \frac{1}{nK} \left( \sum_{k=1}^{K} d_H(X_k, \widehat{X}_{\sigma(k)}) \right)\,,
\end{equation}
tells us, on average, the proportion of bits that differ between our estimated binary vectors and the true ones. 

\begin{table}[h!]
\centering
\caption{BER of {\EM} Algorithm for $K = 8$, Depth = 800}
\label{tab:ber_k8_depth800}
\renewcommand{\arraystretch}{1.3}
\setlength{\tabcolsep}{20pt}
{\fontsize{10pt}{12pt}\selectfont
\begin{tabular}{|c|c|}
\hline
\textbf{Number of qubits} & \textbf{Bit Error Rate} \\ \hline
10         & 0.003      \\ \hline
12         & 0.007      \\ \hline
14         & 0.000          \\ \hline
\end{tabular}
}
\end{table}

Estimating the number of solutions, $K$, is also imperfect. The procedure does occasionally fail to determine the correct number of $K$ solutions. We compute the rate of incorrectly estimating $K$,
\begin{equation}\label{incorrect_K}
P_{K \text{error}} = \frac{N_{\text{incorrectly estimated } K}}{N_{\text{total}}}\,,
\end{equation}
as a simple proportion of number of times the {\EM} algorithm incorrectly estimates $K$, $N_{\text{incorrectly estimated } K}$, to the total number of time the {\EM} algorithm is executed, $N_{\text{total}}$.

Statistical error mitigation approaches often use Hellinger fidelity to gauge accuracy~\cite{hammer,qbeep}. We also adopt this metric to compare our results directly with existing schemes. The Hellinger fidelity, 
\begin{equation}\label{hellinger}
    H_F(P,Q)= \left(\sum_1^L \sqrt{p_i q_i}\right)^2\,,
\end{equation}
measures the difference between two discrete probability distributions, $P$ and $Q$, to quantify their similarity.

\subsection{Qiskit simulations}
\subsubsection{Methodology}
Simulated quantum results are generated for random circuits with $K=\{2,4,6,8\}$ valid noiseless outputs, $n = \{10,12,14\}$ qubits, and circuit depths $D=\{500,600,700,800\}$. Importantly, $K$ is finite for all simulations, including those using other {\qem} algorithms and implicitly excludes algorithms with non-finite $K$, such as variational algorithms. For each configuration we define a target state by randomly selecting $K$ solution vectors from the space of the $2^n$ possible vectors and assigning them equal amplitudes. We then initialize the circuit state vector to this target state using the Qiskit \textit{Initialize} class \cite{qiskit2024}, after which we apply a random entangling circuit and its inverse to reach the desired circuit depth $D$. Without noise, this sequence would output the prepared state. Noise is applied to every gate in the circuit using Qiskit's AerSimulator with a noise model loaded from the IBM Brisbame backend, introducing probabilistic error without altering the intended output state. Each simulation runs with $S = 10{,}000$ shots. After measurement, the circuit collapses into a classical bit-string, which is recorded for each shot and used as input to the {\EM} algorithm.

Algorithm~\ref{alg:junk_shots_filter} filters the data to remove most of the depolarization noise. Under each set of parameters, the {\EM} algorithm does not know the number of valid noiseless solutions, $K$, or the depth, $D$. We give the {\EM} algorithm the data set post the filtering operation. Algorithm~\ref{alg:EM} is the {\EM} algorithm that then generates an {\ml} solution based on the data set. The estimated set of solution vectors, $\widehat{\bX}$, has initial values chosen using the \textit{k-means++} algorithm~\cite{arthur2006k}. The corresponding mixture probabilities, $\widehat{\alpha_k}$, are initialized to $\frac{1}{K_\text{max}}$. The bit-flip probabilities, $\widehat{\bvarepsilon}$, are each initialized to 0.25. Each combination of parameters $n$, $k$, and $D$ has 20 iterations and the results are averaged.

\subsubsection{Numerical results}
In all experimental results, we discuss trends in terms of the {\ber}~\eqref{bit_error_rate}. The majority of errors occur because $K$ is incorrectly estimated. Our reported {\ber} uses results where $K$ is correctly estimated. Across all parameter combinations, we recorded a total $P_{K \text{error}}$ of $0.83\%$.

For all parameter configurations with $K < 8$ and circuit depth $D < 800$, the {\EM} algorithm perfectly estimates all the correct outputs \textbf{with a {\ber} of 0}. Table~\ref{tab:ber_k8_depth800} shows the variation of {\ber} with the number of qubits, $n$, when $K = 8$ and $D = 800$.

\begin{figure}[h]
    \centering
    \includegraphics[width=\linewidth]{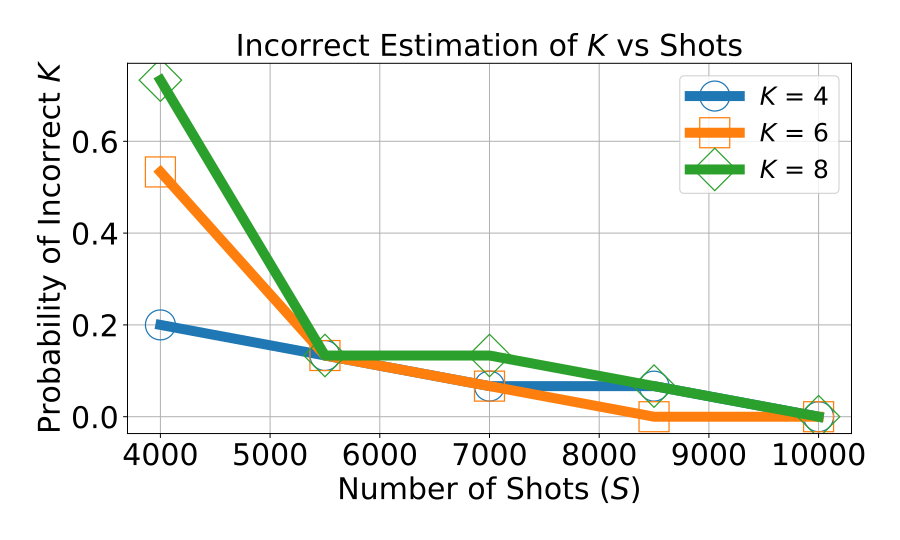}
    \caption{\textbf{Evolution of incorrect $K$ estimation probability with increaing number of shots}. From the simulated 10,000 shots, we use random sampling of the 10,000 shots to determine how $P_{K \text{error}}$ evolves with the number of shots. The plot displays the value of $P_{K \text{error}}$ at a circuit depth of 800, averaged across $n$ for multiple values of $K$. The plot shows a clear increase in $P_{K \text{error}}$ as the number of shots decreases.}
    \label{fig:incorrect_K_shots}
\end{figure}

We further analyze the relationship between $P_{K \text{error}}$ and the number of shots used in the {\EM} algorithm. Depicted in Fig.~\ref{fig:incorrect_K_shots}, $P_{K \text{error}}$ converges to 0 as the number of shots increases for all $K$. Generally, the {\EM} algorithm's performance degrades as $K$ increases. Additionally, for the problem sizes chosen, 10,000 shots appear to be the ideal number where $K$ is estimated perfectly across all configurations.

\begin{table}[h!]
\centering
\caption{Comparison Between QEM Methods}
\label{tab:fidelity_comparison}
\renewcommand{\arraystretch}{1.3}
\setlength{\tabcolsep}{4pt} % reduced column spacing
{\fontsize{9.5pt}{12pt}\selectfont
\begin{tabular}{|c|c|c|c|c|}
\hline
\textbf{Circuit} & \textbf{HAMMER~\cite{hammer}} & \textbf{M3~\cite{M3}} & \textbf{Q-BEEP~\cite{qbeep}} & \textbf{EM} \\ \hline
bv\_n14      & 0.1106 & 0.0219 & 0.0194 & 1.000  \\ \hline
ghz\_n11     & 0.8505 & 0.4831 & 0.4043 & 0.998  \\ \hline
w\_n3        & 0.8902 & 0.9230 & 0.8715 & 1.000  \\ \hline
adder\_n10   & 0.2549 & 0.1628 & 0.1497 & 1.000 \\ \hline
\end{tabular}
}
\end{table}

\subsection{Comparison with other QEM approaches on IBM hardware}

To verify our results, we compare the fidelity using the HAMMER~\cite{hammer}, M3~\cite{M3}, and Q-BEEP~\cite{qbeep} mitigation schemes on a subset of the QASMBench benchmark circuits that were run on the IBM Brisbane QPU. Table~\ref{tab:fidelity_comparison} shows the fidelities of the mitigated probability distributions from the Bernstein-Vazirani, GHZ state, W state, cat state, and adder circuits ranging from 3 to 14 qubits. We demonstrate marked improvements in fidelity and can mitigate outputs well even at higher qubit counts. This is, however, for a limited problem set in which there are only $K$ ground-truth solutions to a quantum algorithm.

\subsection{Scaling with more qubits}
An ideal QEM scheme must be effective as the number of qubits $n$ increases. To demonstrate scalability of the EM algorithm, we test $n=128$ qubits with $K={2,4,8}$ solutions. Simulating such large circuits is infeasible, so we generate synthetic results. With $S=20,\!000$ shots, depolarizing noise ($p=0.9$) corrupts the solution vector, followed by per bit bit-flip with probability $\varepsilon_j$ selected uniformly randomly from $[0.05, 0.15]$. This models both depolarizing noise and bit-level corruption from gate or readout errors. The EM algorithm is run on these noisy outputs, with each $K$ repeated 10 times. In every case, the BER was 0 for all binary vectors. These results strongly suggest that the EM algorithm scales well with large $n$ and achieves high estimation accuracy.

\section{Conclusion}
As we work toward fault-tolerant quantum machines, noisy intermediate-scale quantum (NISQ)-era strategies like quantum error mitigation ({\qem}) are vital for producing reliable computational results. Our statistical approach to {\qem} addresses shortcomings in quantum computation by estimating the most likely noiseless outputs from noisy data. We mitigate the impact of depolarizing noise through filtering, then apply the expectation-maximization ({\EM}) algorithm to compute a maximum likelihood ({\ml}) estimate of the quantum state's distribution in the computational basis. Our {\EM}-based method succeeds for $n = 10, 12, 14$ qubits with $K = 2, 4, 6, 8$ valid solutions, under calibrated machine noise models. In this small-qubit regime, it outperforms other statistical QEM methods such as HAMMER, M3, and Q-BEEP~\cite{hammer,M3,qbeep}. We further scale to larger $n$ using synthetically generated data consistent with depolarizing and bit-flip noise, demonstrating that the {\EM} algorithm retains high accuracy and scalability.

While effective, our approach has limitations. We assume symmetric bit-flip probabilities between 0 and 1, whereas real hardware may exhibit asymmetry, and we do not model complex readout errors such as qubit-to-qubit crosstalk. The finite-$K$ assumption is also limiting, leaving extension to variational algorithms as future work.

The {\EM} algorithm implementation is also suboptimal: initialization heavily influences {\ml} estimation, and estimation of $K$ can occasionally fail. Our filtering strategy for depolarizing noise is heuristic and warrants more rigorous analysis. These challenges offer several directions for future work.

To conclude, we have demonstrated that the {\EM} algorithm offers a promising and scalable framework for quantum error mitigation. Even with a non-ideal implementation, our results highlight its effectiveness and potential for further advances.

\section*{Acknowledgment}

We thank Hrushikesh Pramod Patil and Huiyang Zhou for numerous conversations about QEM.
We also thank Mohsen Ghodrati for his helpful comments about the filters used here. Finally, we thank the anonymous IEEE QCE 2025 reviewers whose comments helped us greatly improve the clarity of the manuscript.

\bibliographystyle{IEEEtran}
\bibliography{bibliography_ieee_refined}

\begin{thebibliography}{10}
\providecommand{\url}[1]{#1}
\csname url@samestyle\endcsname
\providecommand{\newblock}{\relax}
\providecommand{\bibinfo}[2]{#2}
\providecommand{\BIBentrySTDinterwordspacing}{\spaceskip=0pt\relax}
\providecommand{\BIBentryALTinterwordstretchfactor}{4}
\providecommand{\BIBentryALTinterwordspacing}{\spaceskip=\fontdimen2\font plus
\BIBentryALTinterwordstretchfactor\fontdimen3\font minus \fontdimen4\font\relax}
\providecommand{\BIBforeignlanguage}[2]{{%
\expandafter\ifx\csname l@#1\endcsname\relax
\typeout{** WARNING: IEEEtran.bst: No hyphenation pattern has been}%
\typeout{** loaded for the language `#1'. Using the pattern for}%
\typeout{** the default language instead.}%
\else
\language=\csname l@#1\endcsname
\fi
#2}}
\providecommand{\BIBdecl}{\relax}
\BIBdecl

\bibitem{NISQ}
M.~Brooks, ``Beyond quantum supremacy: {T}he hunt for useful quantum computers,'' \emph{Nature}, vol. 574, pp. 19--21, 10 2019.

\bibitem{QEM}
Z.~Cai, R.~Babbush, S.~C. Benjamin, S.~Endo, W.~J. Huggins, Y.~Li, J.~R. McClean, and T.~E. O’Brien, ``Quantum error mitigation,'' \emph{Rev. Mod. Phys.}, vol.~95, no.~4, Dec. 2023.

\bibitem{Temme_EM}
K.~Temme, S.~Bravyi, and J.~M. Gambetta, ``Error mitigation for short-depth quantum circuits,'' \emph{Phys. Rev. Lett.}, vol. 119, p. 180509, Nov 2017.

\bibitem{Endo_EM}
S.~Endo, S.~C. Benjamin, and Y.~Li, ``Practical quantum error mitigation for near-future applications,'' \emph{Phys. Rev. X}, vol.~8, p. 031027, Jul 2018.

\bibitem{Song_2019}
C.~Song \emph{et~al.}, ``Quantum computation with universal error mitigation on a superconducting quantum processor,'' \emph{Science Advances}, vol.~5, no.~9, sep 2019.

\bibitem{Quek_Bounds}
Y.~Quek \emph{et~al.}, ``Exponentially tighter bounds on limitations of quantum error mitigation,'' \emph{Nat. Phys.}, vol.~20, no.~10, pp. 1648--1658, Oct 2024.

\bibitem{Clifford_EM}
P.~Czarnik, A.~Arrasmith, P.~J. Coles, and L.~Cincio, ``Error mitigation with clifford quantum-circuit data,'' \emph{{Quantum}}, vol.~5, p. 592, nov 2021.

\bibitem{bayesian_unfolding}
B.~Pokharel, S.~Srinivasan, G.~Quiroz, and B.~Boots, ``Scalable measurement error mitigation via iterative {Bayesian} unfolding,'' \emph{Phys. Rev. Res.}, vol.~6, p. 013187, Feb 2024.

\bibitem{QMV}
D.~Baron, H.~P. Patil, and H.~Zhou, ``Qubit-wise majority vote: Maximum likelihood quantum error mitigation for algorithms with a single correct output,'' in \emph{Proc. 2024 IEEE Int. Conf. Quantum Comput. Eng.}, vol.~01, 2024, pp. 124--133.

\bibitem{hammer}
S.~Tannu, P.~Das, R.~Ayanzadeh, and M.~Qureshi, ``{H}ammer: Boosting fidelity of noisy quantum circuits by exploiting {H}amming behavior of erroneous outcomes,'' in \emph{Proc. 27th ACM Int. Conf. Archit. Support Program. Lang. Oper. Syst. (ASPLOS)}, 2022, pp. 529--540.

\bibitem{qbeep}
S.~Stein, N.~Wiebe, Y.~Ding, J.~Ang, and A.~Li, ``Q-beep: Quantum bayesian error mitigation employing poisson modeling over the hamming spectrum,'' in \emph{Proceedings of the 50th Annual International Symposium on Computer Architecture}, ser. ISCA '23.\hskip 1em plus 0.5em minus 0.4em\relax New York, NY, USA: Association for Computing Machinery, 2023.

\bibitem{M3}
P.~D. Nation, H.~Kang, N.~Sundaresan, and J.~M. Gambetta, ``Scalable mitigation of measurement errors on quantum computers,'' \emph{PRX Quantum}, vol.~2, no.~4, p. 040326, 2021.

\bibitem{VanTrees}
H.~Van~Trees, \emph{Detection, Estimation, and Modulation Theory}.\hskip 1em plus 0.5em minus 0.4em\relax Wiley, 1968, no. pt. 1.

\bibitem{prorakis}
J.~G. Proakis and D.~G. Manolakis, \emph{Digital signal processing (3rd ed.): principles, algorithms, and applications}.\hskip 1em plus 0.5em minus 0.4em\relax USA: Prentice-Hall, Inc., 1996.

\bibitem{EM_original}
A.~P. Dempster, N.~M. Laird, and D.~B. Rubin, ``Maximum likelihood from incomplete data via the em algorithm,'' \emph{J. R. Stat. Soc., Ser. B (Methodol.)}, vol.~39, no.~1, pp. 1--22, 12 2018.

\bibitem{Mario_K}
M.~Figueiredo and A.~Jain, ``Unsupervised learning of finite mixture models,'' \emph{IEEE Trans. Pattern Anal. Mach. Intell.}, vol.~24, no.~3, pp. 381--396, 2002.

\bibitem{Wilde}
M.~M. Wilde, \emph{Quantum Information Theory}.\hskip 1em plus 0.5em minus 0.4em\relax Cambridge University Press, 2013.

\bibitem{Yang}
J.~Yang, S.~Rahardja, and P.~Fränti, ``Mean-shift outlier detection and filtering,'' \emph{Pattern Recognition}, vol. 115, p. 107874, 2021.

\bibitem{Ramaswamy}
S.~Ramaswamy, R.~Rastogi, and K.~Shim, ``Efficient algorithms for mining outliers from large data sets,'' in \emph{Proc. 2000 ACM SIGMOD Int. Conf. Manage. Data}, ser. SIGMOD '00.\hskip 1em plus 0.5em minus 0.4em\relax New York, NY, USA: Association for Computing Machinery, 2000, p. 427–438.

\bibitem{DBSCAN}
M.~Ester, H.-P. Kriegel, J.~Sander, and X.~Xu, ``A density-based algorithm for discovering clusters in large spatial databases with noise,'' in \emph{Proc. 2nd Int. Conf. Knowl. Discov. Data Min.}, ser. KDD'96.\hskip 1em plus 0.5em minus 0.4em\relax AAAI Press, 1996, p. 226–231.

\bibitem{Bernstein}
E.~Bernstein and U.~Vazirani, ``Quantum complexity theory,'' \emph{SIAM J. Comput.}, vol.~26, no.~5, pp. 1411--1473, 1997.

\bibitem{urbanek2021mitig}
M.~Urbanek, B.~Nachman, V.~R. Pascuzzi, A.~He, C.~W. Bauer, and W.~A. de~Jong, ``Mitigating depolarizing noise on quantum computers with noise-estimation circuits,'' \emph{Physical review letters}, vol. 127, no.~27, p. 270502, 2021.

\bibitem{bennett1996mixed}
C.~H. Bennett, D.~P. DiVincenzo, J.~A. Smolin, and W.~K. Wootters, ``Mixed-state entanglement and quantum error correction,'' \emph{Physical Review A}, vol.~54, no.~5, p. 3824, 1996.

\bibitem{cai2019constructing}
Z.~Cai and S.~C. Benjamin, ``Constructing smaller pauli twirling sets for arbitrary error channels,'' \emph{Scientific reports}, vol.~9, no.~1, p. 11281, 2019.

\bibitem{qiskit2024}
Javadi-Abhari \emph{et~al.}, ``Quantum computing with {Q}iskit,'' 2024.

\bibitem{arthur2006k}
D.~Arthur and S.~Vassilvitskii, ``k-means++: The advantages of careful seeding,'' Stanford, Tech. Rep., 2006.

\end{thebibliography}

\end{document}